# Intelligent System for Speaker Identification using Lip features with PCA and ICA

Anuj Mehra    Anupam Shukla   Mahender Kumawat    Rajiv Ranjan    Ritu Tiwari

**Abstract**—Biometric authentication techniques are more consistent and efficient than conventional authentication techniques and can be used in monitoring, transaction authentication, information retrieval, access control, forensics, etc. In this paper, we have presented a detailed comparative analysis between Principle Component Analysis (PCA) and Independent Component Analysis (ICA) which are used for feature extraction on the basis of different Artificial Neural Network (ANN) such as Back Propagation (BP), Radial Basis Function (RBF) and Learning Vector Quantization (LVQ). In this paper, we have chosen "TULIPS1 database, (Movellan, 1995)" which is a small audiovisual database of 12 subjects saying the first 4 digits in English for the incorporation of above methods. The six geometric lip features i.e. height of the outer corners of the mouth, width of the outer corners of the mouth, height of the inner corners of the mouth, width of the inner corners of the mouth, height of the upper lip, and height of the lower lip which extracts the identity relevant information are considered for the research work. After the comprehensive analysis and evaluation a maximum of 91.07% accuracy in speaker recognition is achieved using PCA and RBF and 87.36% accuracy is achieved using ICA and RBF. Speaker identification has a wide scope of applications such as access control, monitoring, transaction authentication, information retrieval, forensics, etc.

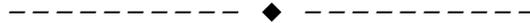

──────── ◆ ─────────

## 1 INTRODUCTION

Speaker Identification can be introduced to make system/services accessed only by the authorized personnel. The possible applications could vary from retrieval of private information to control of financial transactions, access for the authorized persons in a building and so on [2]. User could be authenticated by old conventional techniques (password/personal identification number, token/card) or by its own biometric characteristic (lips, face, iris, fingerprint, voice, etc.). Previous researchers have proposed different algorithms that automatically extract lip features from speaker images [9]. Previous researches have also shown the different proposed techniques using lips feature to authenticate the speaker. Previous work has been done on the incorporation of Active Shape Model (ASM) [7] features and Hidden Markov Model (HMM). The approaches mentioned above are Brown et al. [3] which extracts the identity relevant information by the usage of six geometric lip features and two inner mouth features, Wark and Sridharan [4] that combines the normal from profiles to the contour points to form the grand profile vector (GPV) for the image, Luettin et al. [5] which extracts the Active Shape Model (ASM) features, and uses ASM to model the lip shape and the intensity profile vector along normal to the model points, Matthews et al. [6] which were proposed due to the intensity variation inside the outer lip contour. The main aim of this research is to make a comparative analysis between the techniques PCA and ICA which are used for extraction of lip features on the basis of BP, LVQ, and RBF. In this paper; we have used the six geometric lip features used for the speaker identification. PCA is an approach which uses lower dimensional feature vectors to approximate the original data. PCA attempts to minimize the reconstruction error under the restrictions such as linear reconstruction, and orthogonal factors and maximize the variance [12, 15]. ICA is a technique that seeks statistically independent and non-Gaussian components to find the multi-dimensional data and their underlying factors. It is basically the recognition & separation of mixtures of sources with little prior information. [17]. ICA can be used in a diversified range of applications such as Medical signal processing, Feature extraction, face recognition, Watermarking, Clustering, Time series analysis, and many more applications [16]. Previous researches have been accomplished by integrating Hidden Markov Model (HMM) [8], as compared to our research which uses PCA [12, 13] and ICA [16, 17] for feature extraction and BP, RBF and LVQ as neural network classifiers. Section 2 explains the methodology used in the research whereas Section 3 explains the experiments and the results, and Section 4 elucidate about Conclusion and the future scope. The block diagram of speaker identification process in this work is shown in Fig. 1.

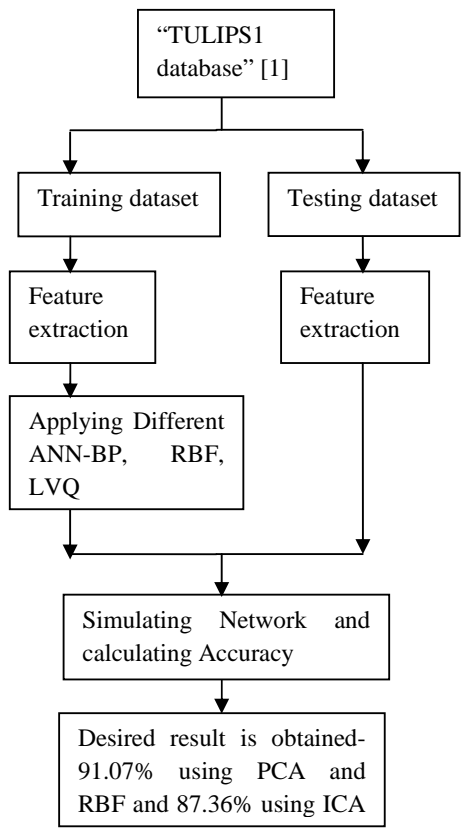

Fig. 1. Block Diagram for Speaker Identification.

## 2 METHODOLOGY

Both PCA and ICA are used as a preprocessing technique for the lip images. PCA is a technique which produces feature vectors in a reduced dimension whereas ICA is a technique that seeks statistically independent and non-Gaussian components to find the multi-dimensional data and their underlying factors. The feature vectors obtained in both the above mentioned processes are then used as an input data to the neural network classifiers i.e. Back Propagation Multi-layer neural network structure (BP-MLNN), Radial basis function (RBF) and Learning Vector quantization (LVQ).

### 2.1. PCA PREPROCESSING

PCA provides an effective technique for dimensionality reduction [12, 15]. The procedure is to first compute the eigen values and eigenvectors from the covariance matrix of the original input data vector. It then computes the orthonormal vectors which are the basis of the computed data. These are the unit vectors and are referred to as the Principle Components. These principle components serve as the new set of axes for the input data.

Let $X = (x_1, x_2, \ldots, x_i, \ldots, x_n)$ represents the $n \times N$ data matrix, where each $x_i$ is a lip vector of dimension $n$, concatenated from a $p \times q$ *lip* image.

$$Y = W^T q \qquad (1)$$

Where Y is the $m \times N$ feature vector matrix, $m$ is the dimension of the feature vector and transformation matrix $W$ is an $n \times m$ transformation matrix whose columns are the eigenvectors corresponding to the $m$ largest eigen values computed according to the formula

The principle component $w_1$ of the data set X is as follows:

$$w_1 = \arg\max_{\|w\|=1} \text{var}\{W^T x\} \qquad (2)$$

### 2.2. ICA PREPROCESSING

Independent component analysis (ICA) [16, 17] is a method that seeks statistically independent and non-Gaussian components which makes it different from the other feature extraction methods. It is a method to find the multivariate or multi-dimensional data and their underlying factors. A broad idea of ICA is shown in Fig. 2.

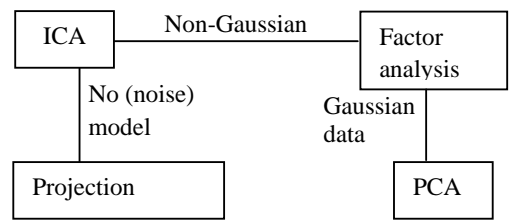

Fig. 2. ICA model and related methods.

ICA model can be explained in the given below equation,

$$X = AS \qquad (3)$$

where, x and s be two random vectors representing $x_1, s_2, x_3, \ldots x_n$ and $s_1, s_2, s_3, \ldots s_n$ respectively and A represents the linear static transformation with elements $a_{ij}$. Thus, $x^T$ is a row vector. The above equation can be rewritten as

$$x = \sum_{i=1}^{n} a_i s_i \qquad (4)$$

It is also known as blind source separation. For simplicity, we have used FastICA algorithm and have converted the noisy database into noise-free by the whitening process as it is useful for our research work.

## 2.3. BACK PROPAGATION MULTI-LAYER NEURAL NETWORKS STRUCTURE

In back propagation the partial derivatives of the cost function with respect to the free parameters of the network are being determined by back propagation error signals through the network, layer by layer. The feed forward multi-layer neural networks are used as a classifier for the PCA output data. The employed neural network is a feed forward multilayer neural network hidden layer as given in the following Fig.3.

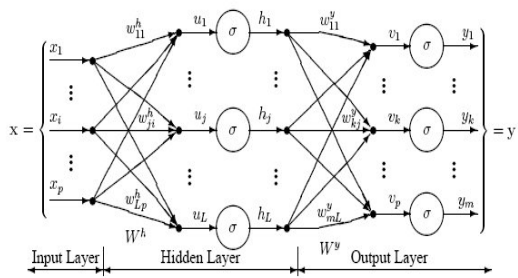

Fig. 3. Back Propagation Multi-layer neural network structure.

In Fig. 3, we have considered a training data which consists of the pairs (x,y) where vector x represents a pattern of input to the network, and the vector y the corresponding desired output (target).According to the error back propagation algorithm, the gradient $\Delta w_{ij}$ can be computed by,

$$\Delta w_{ij} = -\frac{\partial E}{\partial net_i}\frac{\partial net_i}{\partial w_{ij}} \qquad (5)$$

where, weight from unit j to unit i is represented by $w_{ij}$, E is the output error, first factor represents the error of unit i, and the second factor represents the activity of the output unit $y_j$. The activity of the input units $y_i$ is calculated by the network's external input x.

$$y_i = f_i(\sum_{j \in A_i} w_{ij} y_j) \qquad (6)$$

where, $A_i$ is the activity of all the anterior nodes.For hidden units, as the name suggests we must propagate the error back from the output nodes; equation (7) gives us the equation of error back propagation

$$\delta j = -\sum_{i \in P_j} \frac{\partial E}{\partial net_i}\frac{\partial net_i}{\partial y_j}\frac{\partial y_j}{\partial net_j} \qquad (7)$$

where, first factor represents the error of node i, second factor is the weight from unit j to i, and third factor is the derivative of node j's activation function.
Back Propagation networks frequently use one or more hidden layers of sigmoid neurons and output layer of linear neurons [12].

## 2.4. RADIAL BASIS FUNCTION NETWORKS

These networks find the input to output map using local approximators and are extremely fast and require fewer training samples. It has a static Gaussian function as the nonlinearity for the hidden layer processing elements which strictly responds to a small region of the input space where the Gaussian is centered. An unsupervised approach to find the centre of Gaussian function instead of the supervised approach leads us to the better results.

The simulation begins with the training of the unsupervised layer which derives the Gaussian centres and the width from the input data. The output layer is being derived from the input data weighted by a given Gaussian mixture. As the training is completed the supervised segments sets the centres of the Gaussian function and determines the width of each Gaussian. A typical RBF [13, 14] neural network structure is shown in Fig. 4.

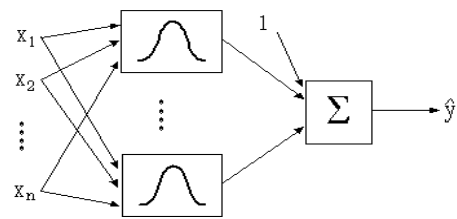

Fig. 4. RBF network structure with one output.

In the network as shown in Fig. 4, $n$ – dimensional input feature vector are accepted to the input layer which is a set of $n$ units consisting of input vectors $x_1, x_2, \ldots, x_n$ and are input to the $l$ hidden functions. The output of the hidden function is then multiplied by the weighting factor $w(i, j)$ which is the input to the output layer of the network $y(x)$.

$$y(x) = \sum_{i=1}^{N} w_i \Phi(\| x - c_i \|) \qquad (8)$$

where the approximating function $y(x)$ is represented as a sum of $N$ radial basis functions, each associated with a different centre $c_i$, and weighted by an appropriate coefficient $w_i$, and $\| \; \|$ indicates the Euclidean norm on the input space.

## 2.5. LEARNING VECTOR QUANTIZATION NETWORKS

LVQ is a prototype-based supervised classification algorithm. LVQ [10] is a supervised learning technique that uses class information to move Voronoi vectors [13] to some extent, so as to improve the quality of classifier decision regions.

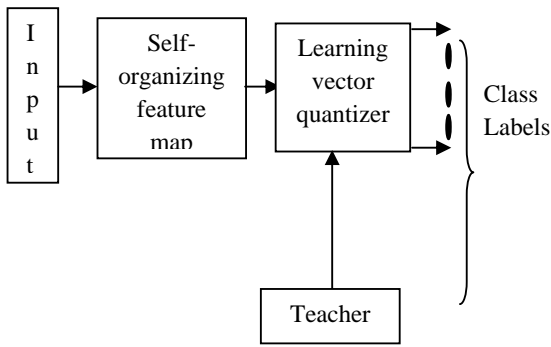

Fig. 5. Block Diagram of Pattern Classification, using a self-feature organizing map and LVQ.

In Fig. 5, it has been shown that to solve a pattern classification problem, the computation of the feature map may be viewed in two stages. The first stage is self-organizing feature map and the second stage is LVQ.

According to LVQ algorithm [11],
(a) If $\zeta_{w_c} = \zeta_{x_i}$

$$w_c(n+1) = w_c(n) + \alpha_n [x_i - w_c(n)]$$

where, $0 < \alpha_n < 1$. (9)

else $\zeta_{w_c} \neq \zeta_{x_i}$

$$w_c(n+1) = w_c(n) - \alpha_n [x_i - w_c(n)] \qquad (10)$$

where, $\{W_j\}_{j=1}^{l}$ denote the set of Voronoi vectors, $\{X_i\}_{i=1}^{N}$ denote the set of input vectors $\zeta_{w_c}$ denote the class associated with the Voronoi vector, $\zeta_{x_i}$ denote the class label of the input vector $x_i$, and $\alpha_n$ represents the learning constant with the number of iterations n.
(b) The other Voronoi vectors are not modified.

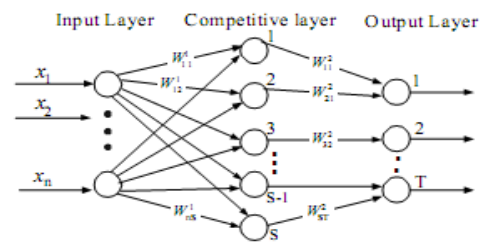

Fig. 6. Architecture of an LVQ neural network.

As shown in Fig. 6, an LVQ network has a first competitive layer and a second linear layer. We have considered a set of n input vectors $x_1, x_2, \ldots, x_n$ which are multiplied to the weights $w_{ij}$ to compute the output. The weight vector of the competitive layer is shown by the equation (10) where $w_c(n+1)$ the adjusted weight vector is after adjustment.

## 3 EXPERIMENT AND RESULTS

In this research work, we study and explore the use of neural networks in the speaker using lip features. First of all, the database named "TULIPS1 database, (Movellan, 1995)" (shown in Fig. 7) is selected. The database consists of 12 subjects saying the first 4 digits in English and the digits are repeated twice by each speaker. Each digit spoken constitutes of 6 instances and the size of each image is 100 by 75 pixels. In the second step, the database is split into two datasets i.e. training and testing dataset. In this experiment, 7 subjects are chosen for the research work constituting in the formulation of 7 classes with 24 images present in each class i.e. 168 images are selected for training and 168 images are selected for testing dataset. In the next step,

PCA and ICA are applied on training dataset for feature extraction followed by the neural network classifiers (BP, RBF, and LVQ). An output matrix is achieved which consists of the reduced set of images. A target matrix is formed and testing images are trained using PCA and ICA followed by the simulation of the neural networks. In the last step, accuracy is calculated by the number of images of training and testing dataset matched.

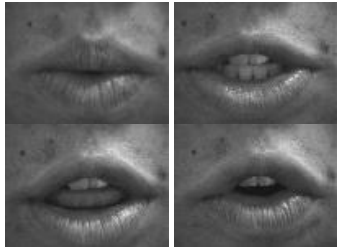

Fig. 7. Speaker speaking one to four

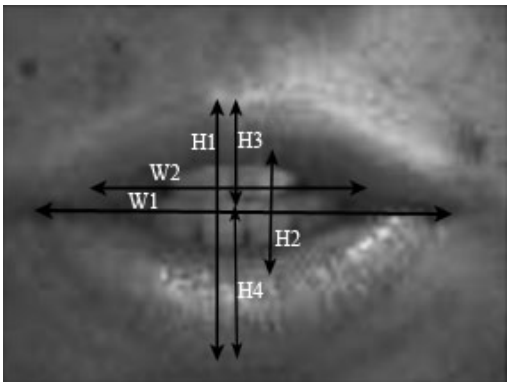

Fig. 8. Image showing lip features

As shown in Fig. 8, H1 represents height of the outer corners of the mouth, W1 represents width of the outer corners of the mouth, H2 represents height of the inner corners of the mouth, W2 represents width of the inner corners of the mouth, H3 represents height of the upper lip, and H4 represents height of the lower lip.

### 3.1. CLASSIFICATION BY THE BACK PROPAGATION NETWORK

PCA and ICA along with BP is applied to these lip images for training and then it is tested with rest of the images. Here, two hidden layers are used in BP.

Table 1
Statistical Data of BP

|  | PCA | ICA |
|---|---|---|
| Input Vector Nodes | 6 | 6 |
| Number of Hidden Layers | 2 | 2 |
| Number of neurons (hidden1 ,hidden2 & output) | 20,25,7 | 20,25,7 |
| Transfer functions (input , hidden & output ) | Tansigmoid, tansigmoid, linear | Tansigmoid, tansigmoid, linear |
| Network Learning rate | .01 | .01 |

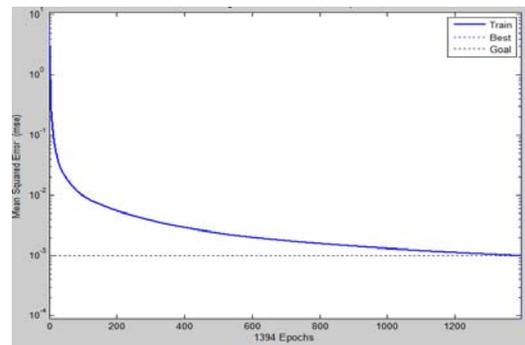

Fig. 9. Learning by PCA along with BP

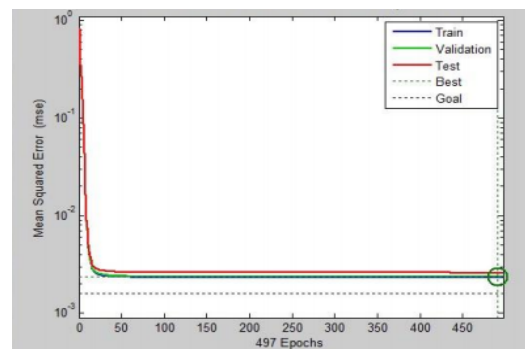

Fig. 10. Learning by ICA along with BP

### 3.2. CLASSIFICATION BY THE RADIAL BASIS FUNCTION NETWORK

PCA and ICA along with RBF is applied to these lip images for training and then it is tested with rest of the images. In RBF a hidden radial basis layer & output linear layer is used.

Table 2
Statistical Data of RBF

|  | PCA | ICA |
|---|---|---|
| Number of Radial Basis Layers | 1 | 1 |
| Number of neurons (input ,radial basis & output) | 6,125,7 | 6,125,7 |
| Spread | 25 | 25 |

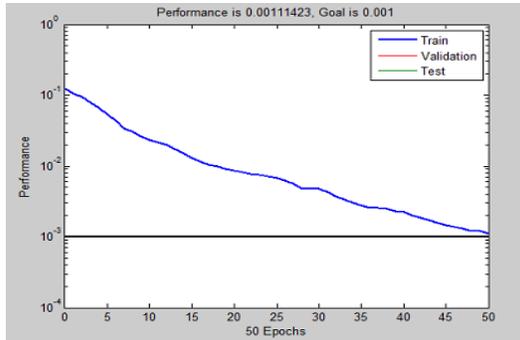

Fig.11. Learning by PCA along with RBF

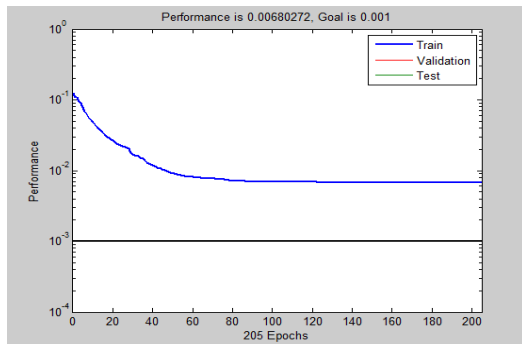

Fig. 12.Learning by ICA along with RBF

### 3.3. CLASSIFICATION BY THE LEARNING VECTOR QUANTIZATION NETWORK

PCA and ICA along with LVQ is applied to these lip images for training and then it is tested with rest of the images. In LVQ a hidden competitive layer and an output linear layer is used.

Table 3
Statistical Data of *LVQ*

|  | PCA | ICA |
|---|---|---|
| Number of competitive Layers | 1 | 1 |
| Number of neurons (input ,competitive & output) | 6,40,7 | 6.40,7 |
| Transfer function | Lvq1.0 | Lvq1.0 |
| Network Learning rate | .001 | .001 |

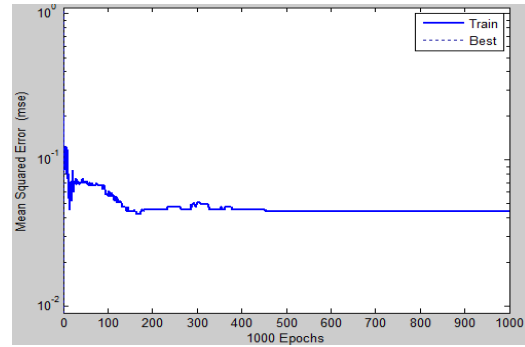

Fig. 13.Learning by PCA along with LVQ

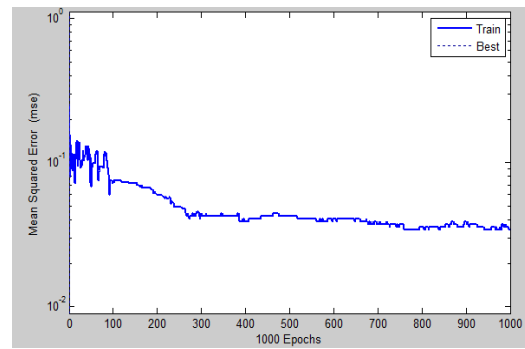

Fig. 14.Learning by ICA along with LVQ

### 3.4. EXPERIMENTAL RESULTS

The results for the recognition test with BP, RBF & LVQ with PCA and ICA are shown in Table IV:

Table 4
Statistical Data of different neural network techniques

| Methods | BP | RBF | LVQ |
|---|---|---|---|
| Recognition Rate with PCA | 89.88% | 91.07% | 87.5% |
| Recognition Rate with ICA | 85.8% | 87.36% | 84.52% |

Result shows that the recognition performance for PCA with RBF is better than the other methods used.

## 4 CONCLUSIONS

It can be concluded from the experimental result section that RBF overrules BP and LVQ when used with PCA and ICA. It can also be concluded that BP network when used either with PCA or ICA achieved better accuracy than LVQ due to its back propagation mechanism. When PCA and ICA are compared with each other it can be seen from the result section that PCA overrules ICA in all the neural network classifiers used. The second conclusion can be made with respect to computational time; PCA took much lesser time than ICA as the former one does not incorporates whitening process as used in the latter one. Also, RBF took lesser computational time than BP and LVQ in both the cases due to the fact that it employs local approximators and hence require fewer training samples. Hence, RBF in lip recognition is fast and robust than the above methods. In future, the work can be done on an identification system that incorporates the features of both lips and speech. Accuracy can be improved by the use of other feature extraction techniques. Accuracy can also be improvised by incorporating several feature extraction techniques to form a unique one.